\documentclass[fleqn,10pt]{wlscirep}
\usepackage[utf8]{inputenc}
\usepackage[T1]{fontenc}
\usepackage{amsthm}
\usepackage{hyperref}
\usepackage{booktabs}
\usepackage{pythonhighlight}
\usepackage{tikz}
\usetikzlibrary{shapes.geometric, arrows}
\usepackage[draft]{minted}
\newenvironment{code}{\captionsetup{type=listing}}{}
\usepackage{caption}
\usepackage{newfloat}

\usepackage{lineno}
\definecolor{light-gray}{gray}{0.95}
\newcommand{\codehighlight}[1]{\colorbox{light-gray}{\small\texttt{#1}}}
\theoremstyle{definition}
\newtheorem{definition}{Definition}
\usepackage{csquotes}
\usepackage[symbol]{footmisc}

\title{An Open Source Python Library for Anonymizing Sensitive Data}

\author[1,*]{Judith Sáinz-Pardo Díaz}
\author[1]{Álvaro López García}

\affil[1]{Instituto de Física de Cantabria (IFCA), CSIC-UC \\ Avda. los Castros s/n. 39005 - Santander (Spain)}

\affil[*]{corresponding author: Judith Sáinz-Pardo Díaz (sainzpardo@ifca.unican.es)}

\begin{abstract}
Open science is a fundamental pillar to promote scientific progress and collaboration, based on the principles of open data, open source and open access. However, the requirements for publishing and sharing open data are in many cases difficult to meet in compliance with strict data protection regulations. Consequently, researchers need to rely on proven methods that allow them to anonymize their data without sharing it with third parties. To this end, this paper presents the implementation of a Python library for the anonymization of sensitive tabular data. This framework provides users with a wide range of anonymization methods that can be applied on the given dataset, including the set of identifiers, quasi-identifiers, generalization hierarchies and allowed level of suppression, along with the sensitive attribute and the level of anonymity required. The library has been implemented following best practices for integration and continuous development, as well as the use of workflows to test code coverage based on unit and functional tests. 
\end{abstract}
\begin{document}

\flushbottom
\maketitle

\thispagestyle{empty}

\vspace{-0.5cm}
\section*{Introduction}\label{sec:intro}

\footnotetext{\textit{Preprint under review.}}
The development of data driven applications is a growing field that requires large volumes of data for its successful evolution and performance. In some cases, these models are powered by data that may contain information about individuals, and therefore it is critical to focus on the privacy of such data. Numerous studies have collected information on the possible bias that algorithms and in particular Artificial Intelligence (AI) models can potentially exhibit based on the data used during training \cite{10.1145/3457607, 10.1145/3631326}. When it comes to data containing personal information, performing proper pre-processing and curation steps are therefore key for two reasons: (1) to protect the privacy of the individuals associated with the data (2) to mitigate the bias that it may contain and that such biases may be propagated in the models. 

Substantial legislative work has been done to advance these technologies with attention to the security of personal data of individuals. Specifically, in Europe the General Data Protection Regulation (GDPR) was put into effect on May 25, 2018, and it aims to protect natural people regarding the processing of their personal data and the free movement of such information within the European Union and the European Economic Area (EEA). The GDPR establishes within its article 5 principles regarding data protection, including that data must be processed in such a way as to ensure adequate security of personal data. Likewise, consideration number 26 of the same states that this regulation does not affect the processing of anonymous information, including for statistical or research purposes, considering as anonymous information that which does not relate to an identified or identifiable natural person \cite{GDPR2016a}. 
Recently, with the rise of technologies based on artificial intelligence and the risks they may entail (among others in relation to security and privacy of the individual) the European Commission proposed on April 21, 2021 to launch the EU AI Act \cite{AIAct}, which was approved on March 13, 2024, composing the first regulation on artificial intelligence. Specifically, this regulation defines different risks related to AI and different levels for them: Unacceptable, high, limited or minimal. As an example, AI systems identified as high-risk include AI technology employed in administration of justice and democratic processes.

Going back to the legislation on privacy and data security, in the USA the Privacy Act \cite{PrivacyAct} (1974) establishes a code of fair information practices that governs the collection, maintenance, use, and dissemination of information about individuals which is maintained by federal agencies. More specifically in the field of health, HIPAA\cite{hipaa} is a suite of federal regulatory standards that outline the use and disclosure of protected health information in the United States.  

In view of the above, it is essential to have tools to ensure privacy through the anonymization of data that may be of a sensitive nature and associated with individuals, especially if there is a potential use of such data through artificial intelligence techniques, more specifically as the basis of data-driven applications, like machine or deep learning (ML/DL) models. To this end, in this article we present \textit{anjana}, an open source Python library that implements different anonymization techniques that can be applied on tabular data without the need to externalize, upload to different service providers or share them with third parties. 

This paper is structured as follows: in the \hyperref[sec:methods]{Methods\ref*{sec:results}} Section we explore state of the art and motivate the need to introduce this open source tool. We also present the anonymity techniques implemented in the library proposed in this study, their connection and some important definitions. In the \hyperref[sec:results]{Results\ref*{sec:results}} Section we explain the functionality of the library and some examples, together with important remarks on the software development. Finally, the \hyperref[sec:discussion]{Discussion\ref*{sec:discussion}} Section provides insights on the management of multiple sensitive attributes, how to create the hierarchies and get the anonymity transformation applied and the incorporation of these anonymity methods in data science workflows, concluding with some directions for future work.

\section*{Methods}\label{sec:methods}

Regarding tabular data anonymization, numerous theoretical papers review the implementation and definition of the techniques that will be implemented in this library and that are reported in the following subsection. In particular, each of these methods each of them focusing on different issues or prevents different attacks, as presented in the diagram of Figure~\ref{fig:diagram}.

Specifically, in 2002 a definition of \textit{k-anonymity} was given by L. Sweeney \cite{sweeney2002k} and in 2006 \textit{($\alpha$,k)-anonymity} was introduced by R. C. Wong et al. in order to ``protect the association of individuals to sensitive information'' \cite{wong2006alpha}. In 2007 A. Machanavajjhala et al. presented \textit{$\ell$-diversity} as a novel approach to protect against some of the attacks to which \textit{k-anonymity} proved to be sensitive \cite{machanavajjhala2007diversity}. In that paper the concepts of \textit{entropy $\ell$-diversity} and \textit{recursive (c,$\ell$)-diversity} are also introduced. In the same year (2007) N. Li et al. presented \textit{t-closeness} as a privacy technique beyond \textit{$\ell$-diversity} and \textit{k-anonymity} \cite{li2006t}. In 2008 J. Brickell and V. Shmatikov \cite{10.1145/1401890.1401904} already define the concept of \textit{$\delta$-disclosure privacy} and they measure the trade-off between privacy and utility using the adult dataset (which is also used for testing purposes in the current work). Finally, in 2012 J. Cao and P. Karras proposed the use \textit{$\beta$-likeness} as a ``robust privacy model for microdata anonymization'', introducing the definitions of both \textit{basic $\beta$-likeness} and \textit{enhanced $\beta$-likeness} \cite{cao2012publishing}.

One of the main concerns arising nowadays in relation to data privacy stems from the important role played by data as the basis for artificial intelligence (AI), machine and deep learning (ML/DL) models \cite{10.1007/978-3-319-45381-1_5}. These kind of models can perform classification, clustering, inference, pattern recognition or anomaly detection tasks based on data that in many cases are subject to privacy restrictions. In the same direction, the development of data-driven AI models is extensively supported in Python by numerous frameworks for both ML and DL, as is the case for \codehighlight{scikit-learn}\cite{sklearn_api}, \codehighlight{keras}\cite{chollet2015keras}, \codehighlight{TensorFlow}\cite{tensorflow2015-whitepaper} and \codehighlight{PyTorch}\cite{NEURIPS2019_9015} among others. However, in relation to the prior anonymization of the data from which the models available in these libraries are based, the computing facilities to apply the above techniques using open source Python frameworks are limited. We can find some projects on GitHub that implement some of the techniques mentioned above, especially in the case of \textit{k-anonymity}, for which we can find different algorithms to implement it, such as \textit{mondrian}, \textit{datafly} or \textit{incognito} \cite{ayala2014systematic, yuvaraj2022privacy}.

As for Python libraries concerning data anonymization, we can highlight \codehighlight{AnonyPy}\cite{anonypy} which implements the \textit{mondrian} algorithm supporting \textit{k-anonymity}, \textit{$\ell$-diversity} and \textit{t-closeness}.
On the other hand, the \codehighlight{anonym}\cite{anonym} library is designed to anonymize dataframes and it operates by replacing real data with fake ones, while maintaining the structure and format of the original data. \codehighlight{dicognito}\cite{dicognito} is a Python library and command line interface (CLI) that anonymizes medical records given in DICOM format by removing or substituting different fields. The \codehighlight{privapy}\cite{privapy} library includes methods for text anonymization by removing identifying or quasi-identifying words, and also provides methods for image cleanup, such as blackout or pixelate. Besides, \codehighlight{python-anonymity} is a master thesis project in which \textit{k-anonymity} was implemented by means of both datafly and incognito, and some preliminary versions for \textit{$\ell$-diversity} and \textit{t-closeness} were provided (this project is read only and archived, since it was created for academic purposes\cite{master-python-anonymity}). 

In terms of open source tools for anonymizing sensitive data, \codehighlight{ARX}\cite{prasser2015putting}, written in Java, stands out. It is a comprehensive open source software for sensitive data anonymization which supports several privacy models (such us the ones exposed in this work), together with risk and quality models. In addition, \codehighlight{ARXaaS}\cite{arxaas} claims to be an ``anonymization as a service'' project built on top of the ARX library, which uses HTTP by default. Additionally, \codehighlight{Amnesia}\cite{amnesia} is a framework deployed in Java for data anonymization that, in addition to \textit{k-anonymity} implements \textit{km-anonymity}, which can provide better assessments in case of high dimensional data. It can be installed locally but also includes an online version that allows to test its functionality. Other software solutions are available, as the ones reviewed in \cite{pycanon_paper}, such as \codehighlight{$\mu$-ARGUS}, \codehighlight{$\tau$-ARGUS}, \codehighlight{PRIVAaaS} or \codehighlight{g9 Anonymize}. Finally, it is important to mention \codehighlight{pycanon} \cite{pycanon_paper}, which will be used as the basis for the library proposed in this paper, and which is an open source Python library that allows to check the anonymity level of a tabular dataset according to the techniques mentioned at the beginning of this section.

Extensive development has taken place in recent years in terms of frameworks for AI/ML/DL data-driven model development built on top of Python. In addition, this review makes it evident that it is necessary to provide users, and in particular researchers working with sensitive tabular data, with a tool that can be easily installed locally and used in an intuitive way to have at their disposal a wide range of anonymization techniques, covering different types of attacks, and that this is done in the form of an open source Python library. This motivates the implementation of the \textit{anjana} open source Python library, which aims to provide a solution to this need. 

\subsection*{Anonymity techniques}

The objective of the open source framework presented in this work is to provide researchers and general users which work with tabular data containing sensitive information, with an easy-to-use Python based tool that allows them to anonymize them according to different techniques.

First of all, let's introduce some basic definitions in relation to the attributes or columns of tabular databases that must also be entered as input to the anonymization techniques implemented. 

\begin{definition}
    Given a tabular database, each column composes an attribute in relation to the data, and they can be classified as follows:  
\end{definition}
    \begin{quote}
        \textit{\textbf{Identifiers (ID)}}, which contain information that allow to unequivocally identify a user (e.g. name, ID card number, national insurance number, etc.). This information must be removed in the anonymized version of the database.\\
        \textit{\textbf{Quasi-identifiers (QI)}}, these are the information that although by themselves do not allow to uniquely identify a user, they do allow to identify him/her through a combination of a series of these (examples: age, city of birth, level of studies, zip code, etc).\\
        \textit{\textbf{Sensitive attributes (SA)}}, which are the information about the individuals in the database that we want to protect and that should not be inferred by an attacker (example: diseases, salary class, medical records, police records, etc).\\
        \textit{\textbf{Insensitive attributes (IN)}}, which do not require special processing and can be left in the same form (example: randomly generated IDs).
    \end{quote}

\begin{definition}
    Given a database $\mathcal{D}$ and a set of \textit{QI}, we define an \textit{\textbf{equivalence class (EC)}} as a subset of rows that are identical regarding the given set of \textit{QI} \cite{pycanon_paper}. 
    
    Formally, be $\mathcal{S} \subseteq \mathcal{D}$ a subset of rows of $\mathcal{D}$, $n_{QI}$ the number of quasi-identifiers in $\mathcal{D}$, $\mathcal{S}$ form an equivalence class if for every pair of entries $x, y \in \mathcal{S}$ and for every quasi-identifier $QI_{i}$ $ i \in \{1,\hdots,n_{QI} \}$, $x[QI_{i}]=y[QI_{i}]$ $\forall i \in \{1,\hdots,n_{QI} \}$ and in addition $\nexists z \in \mathcal{D} \setminus
    \mathcal{S}$ verifying $z[QI_{i}]=x[QI_{i}]$ $\forall i \in \{1,\hdots,n_{QI}\}$ and $x \in \mathcal{S}$.
\end{definition}

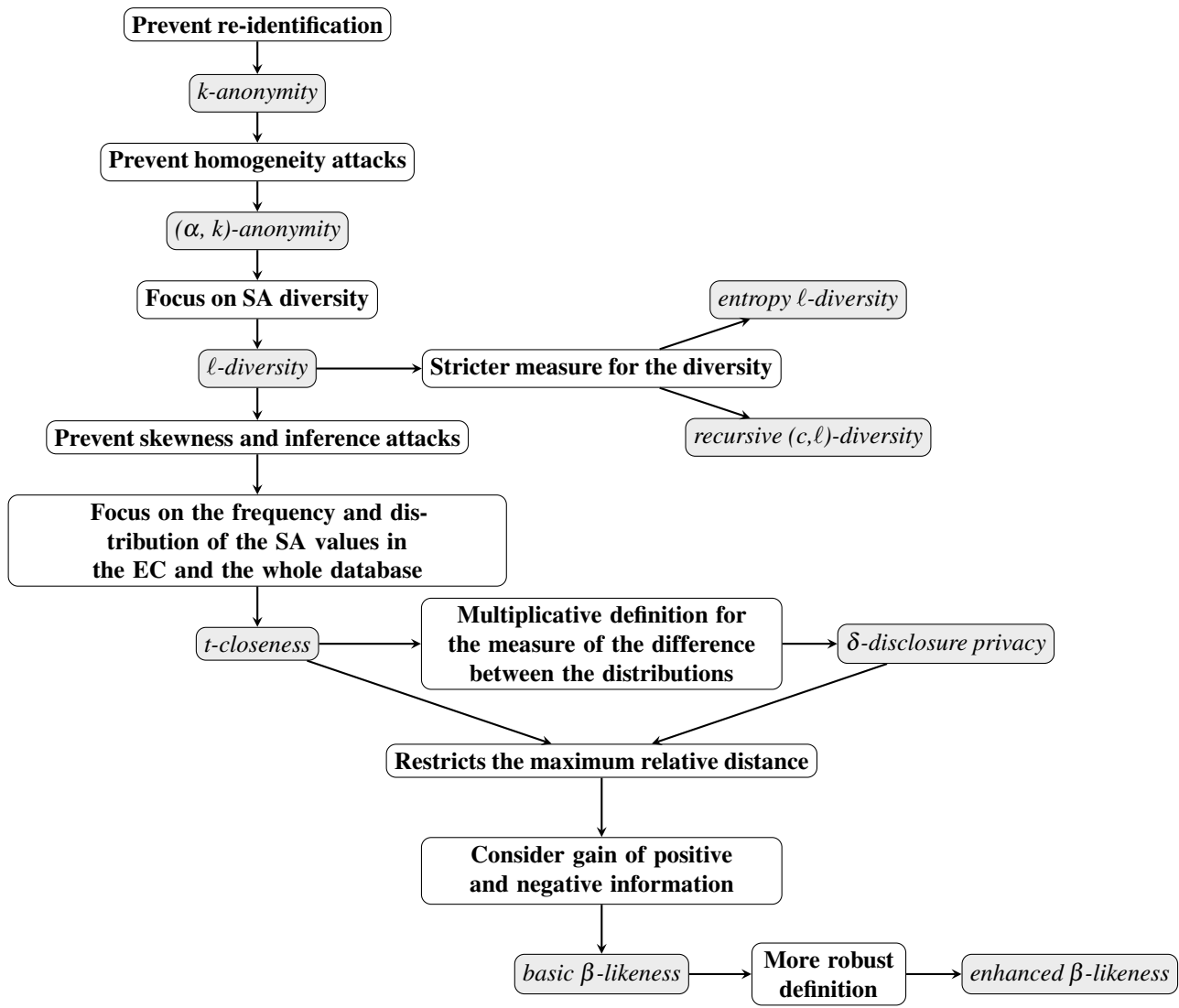
\begin{figure}[ht]
\begin{tikzpicture}
    \node[draw, rectangle, rounded corners] (A) at (0,0) {\textbf{Prevent re-identification}};
    \node[draw,rectangle,rounded corners, fill=gray!15] (B) at (0,-1) {\textit{k-anonymity}};
    \node[draw, rectangle, rounded corners] (C) at (0,-2) {\textbf{Prevent homogeneity attacks}};
    \node[draw,rectangle,rounded corners, fill=gray!15] (D) at (0,-3) {\textit{($\alpha$, k)-anonymity}};
    \node[draw, rectangle, rounded corners] (E) at (0,-4) {\textbf{Focus on SA diversity}};
    \node[draw,rectangle,rounded corners, fill=gray!15] (F) at (0,-5) {\textit{$\ell$-diversity}};
    \node[draw, rectangle, rounded corners] (G) at (5,-5) {\textbf{Stricter measure for the diversity}};
    \node[draw,rectangle,rounded corners, fill=gray!15] (H) at (8,-4) {\textit{entropy $\ell$-diversity}};
    \node[draw,rectangle,rounded corners, fill=gray!15] (I) at (8,-6) {\textit{recursive (c,$\ell$)-diversity}};
    \node[draw, rectangle, rounded corners] (J) at (0,-6) {\textbf{Prevent skewness and inference attacks}};
    \node[draw, rectangle, rounded corners, text width=7cm, text centered] (K) at (0,-7.5) {\textbf{Focus on the frequency and distribution of the SA values in the EC and the whole database}};
    \node[draw,rectangle,rounded corners, fill=gray!15] (L) at (0,-9) {\textit{t-closeness}};
    \node[draw, rectangle, rounded corners, text width=5cm, text centered] (M) at (5,-9) {\textbf{Multiplicative definition for the measure of the difference between the distributions}};
    \node[draw,rectangle,rounded corners, fill=gray!15] (N) at (10,-9) {\textit{$\delta$-disclosure privacy}};
    \node[draw, rectangle, rounded corners, text centered] (O) at (5,-10.7) {\textbf{Restricts the maximum relative distance}};
    \node[draw, rectangle, rounded corners, text width=5cm, text centered] (P) at (5,-12.3) {\textbf{Consider gain of positive and negative information}};
    \node[draw,rectangle,rounded corners, fill=gray!15] (Q) at (5,-13.8) {\textit{basic $\beta$-likeness}};
    \node[draw, rectangle, rounded corners, text width=2cm, text centered] (R) at (8.3,-13.8) {\textbf{More robust definition}};
    \node[draw,rectangle,rounded corners, fill=gray!15] (S) at (11.8,-13.8) {\textit{enhanced $\beta$-likeness}};
    
    \draw[thick,->,>=stealth] (A) -- (B);
    \draw[thick,->,>=stealth] (B) -- (C);
    \draw[thick,->,>=stealth] (C) -- (D);
    \draw[thick,->,>=stealth] (D) -- (E);
    \draw[thick,->,>=stealth] (E) -- (F);
    \draw[thick,->,>=stealth] (F) -- (G);
    \draw[thick,->,>=stealth] (G) -- (H);
    \draw[thick,->,>=stealth] (G) -- (I);
    \draw[thick,->,>=stealth] (F) -- (J);
    \draw[thick,->,>=stealth] (J) -- (K);
    \draw[thick,->,>=stealth] (K) -- (L);
    \draw[thick,->,>=stealth] (L) -- (M);
    \draw[thick,->,>=stealth] (M) -- (N);
    \draw[thick,->,>=stealth] (L) -- (O);
    \draw[thick,->,>=stealth] (N) -- (O);
    \draw[thick,->,>=stealth] (O) -- (P);
    \draw[thick,->,>=stealth] (P) -- (Q);
    \draw[thick,->,>=stealth] (Q) -- (R);
    \draw[thick,->,>=stealth] (R) -- (S);
\end{tikzpicture}
\caption{Workflow: select the anonymity technique to be applied depending on the privacy objective.}
\label{fig:diagram}

\end{figure}

Regarding the anonymity techniques implemented, in the version 1.0.0 of the library, the definition of one sensitive attribute is supported. In this line, \textit{anjana} provides users with the following nine anonymity tools defined as stated in \cite{pycanon_paper} based on the assumption of only one SA: \textit{k-anonymity}, \textit{($\alpha$,k)-anonymity}, \textit{$\ell$-diversity}, \textit{entropy $\ell$-diversity}, \textit{recursive (c,$\ell$)-diversity}, \textit{t-closeness}, \textit{$\delta$-disclosure privacy}, \textit{basic $\beta$-likeness} and \textit{enhanced $\beta$-likeness}. As stated in \cite{pycanon_paper}, these techniques are complementary as they prevent from different types of attacks (see also \cite{prasser2015putting}). We can see the summary of the main attacks prevented in Table 2 from \cite{pycanon_paper}. However, each technique can prevent more attacks than the ones marked in that table, although some are more suitable than others for certain attacks. The workflow that can be followed to choose the appropriate anonymization technique is shown in the diagram presented in Figure~\ref{fig:diagram}. It is important to note that although these functions have been defined for a single sensitive attribute, it is possible to apply them in case of multiple sensitive attributes as will be explained in the \hyperref[sec:discussion]{Discussion\ref*{sec:discussion}} section, following an iterative process depending on the desired paradigm: \textit{harmonization of the quasi-identifiers} or \textit{quasi-identifiers update} \cite{pycanon_paper}. 

The idea of these techniques is to remove the identifiers and apply recursive transformations on the quasi-identifiers that allow them to be generalized, making complicated to extract information about a particular individual. To apply these transformations, a set of hierarchies must be created for each QI. Usually, if we allow the suppression of a quasi-identifier, this will be the last level of hierarchy.

\begin{definition}
    We define a \textit{\textbf{hierarchy}} $h_{1}$ over a quasi-identifier $Q$ of a database $\mathcal{D}$ as a mapping from $D(Q)$ to $D(Q_{1})$, $h_{1}:D(Q)\longrightarrow D(Q_{1})$, with $D(Q)$ the initial set of values of $Q$ in $D$, and $D(Q_{1})$ being the new set of new values of $Q$ once generalized (first transformation). Thus, be $\{x_{1},\hdots,x_{m}\} \in D(Q)$ the set of unique values taken by the attribute $Q$ in $\mathcal{D}$, the function $h_{1}$ assigns $h_{1}(x_{i})=y_{j}$ $\forall i \in \{1,\hdots,m\}$. Note that different values of $x_i$ can take the same value $y_{j}$ $j \in \{1,\hdots, p\}$, $p \leq m$.

    Thus, we can define $n$ mapping $h_{i}$ $\forall i \in \{1,\hdots,n\}$ to generalize the attribute $Q$ as much as necessary by applying a new hierarchy over each state of the database: $h_{1}:D(Q)\longrightarrow D(Q_{1})$, $h_{2}:D(Q_{1})\longrightarrow D(Q_{2})$, $\hdots$, $h_{n}:D(Q_{n-1})\longrightarrow D(Q_{n})$. Then, for the quasi-identifier $Q$ we define a set of hierarchies $H_{Q}=\{h_{1},\hdots,h_{n}\}$, with $n$ the number of possible transformations for $Q$.

    If we perform only the first transformation, $Q$ will have been anonymized with hierarchy level 1, while if we apply $n$ transformations (until $h_n$), we will have a transformation level of $n$ for $Q$
\end{definition}

\section*{Results}\label{sec:results}

\subsection*{Functionality and use examples}
As already stated, \textit{anjana} provides user with the most common anonymization techniques for their application on tabular databases. These methods have been performed according to the definition given in \cite{pycanon_paper}, and the \codehighlight{pycanon} Python library has been used as an auxiliary tool to check that the level of anonymization required by the user is verified.

Except for \textit{k-anonymity}, for which the sensitive attribute is not introduced as it focuses only on the quasi-identifiers, the input of the functions that apply the available anonymity techniques is as follows:

\begin{itemize}
    \item \textbf{Data}: a pandas dataframe with the data to be anonymized. Each column can contain: identifiers, quasi-identifiers or sensitive attributes.
    \item \textbf{Identifiers:} a list of strings containing the names of the identifiers in the dataframe, in order to suppress them (substituting by '*').
    \item \textbf{Quasi-identifiers:} a list of strings, containing the names of the quasi-identifiers in the dataframe.
    \item \textbf{Sensitive-attributes:} a string with the name of the sensitive attribute in the dataset (only one). This parameter must be introduced in case of applying other techniques than \textit{k-anonymity}.
    \item \textbf{Privacy level:} an int or float (depending the function applied) with level of anonymity to be applied, e.g. $k$ (int) for \textit{k-anonymity}, $\alpha$ (float) and $k$ (int) for \textit{($\alpha$,k)-anonymity}, $\ell$ (int) for \textit{$\ell$-diversity} or \textit{entropy $\ell$-diversity}, $c$ (int) and $\ell$ (int) for \textit{recursive (c,$\ell$)-diversity}, $t$ (float) for \textit{t-closeness}, $\beta$ (float) for \textit{basic} or \textit{enhanced $\beta$-likeness} or $\delta$ (float) for \textit{$\delta$-disclosure privacy}. Note that in all the cases a value of $k$ for \textit{k-anonymity} must be introduced, following the same structure as \codehighlight{ARX}. 
    \item \textbf{Suppression level}: a float between 0 and 100 with the maximum level of record suppression allowed.
    \item \textbf{Hierarchies:} dictionary containing the hierarchies and the generalization level for each quasi-identifier to be generalized. The dictionary contains one dictionary for each quasi-identifier with the hierarchies and the levels.
\end{itemize}

In Table~\ref{tab:functions}, the different available techniques are shown, together with their name in the library and the input expected for applying them.

\begin{table}[ht]
    \centering
    \begin{tabular}{rl}
    \toprule
    \textbf{Technique} & \textbf{Function and input}\\
    \midrule
    \textit{k-anonymity} &  \textit{k\_anonymity(data, id, qi, k, supp, hier)}\\
    \textit{($\alpha$,k)-anonymity} & \textit{alpha\_k\_anonymity(data, id, qi, sa, $\alpha$, k, t, supp, hier)}\\
    \textit{$\ell$-diversity} & \textit{l\_diversity(data, id, qi, sa, k, $\ell$, supp, hier)}\\
    \textit{entropy $\ell$-diversity} &  \textit{entropy\_l\_diversity(data, id, qi, sa, k, $\ell$, supp, hier)}\\
    \textit{recursive (c,$\ell$)-diversity} & \textit{recursive\_c\_l\_diversity(data, id, qi, sa, k, c, $\ell$, supp, hier)}\\
    \textit{t-closeness} & \textit{t\_closeness(data, id, qi, sa, k, t, supp, hier)}\\
    \textit{$\delta$-disclosure privacy} & \textit{delta\_disclosure(data, id, qi, sa, k, $\delta$, supp, hier)}\\
    \textit{basic $\beta$-likeness} & \textit{basic\_beta\_likeness(data, id, qi, sa, k, $\beta$, supp, hier)}\\
    \textit{enhanced $\beta$-likeness} & \textit{enhanced\_beta\_likeness(data, id, qi, sa, k, $\beta$, supp, hier)}\\
    \bottomrule
    \end{tabular}
    \caption{\textit{anjana} main functions to apply the different anonymity techniques. Note that \textit{id} is the list of identifiers, \textit{qi} that of the quasi-identifiers, \textit{sa} is the sensitive attribute, \textit{supp} the suppression level, and \textit{hier} the dictionary with the hierarchies. In all the cases $k$ is the desired value for \textit{k-anonymity}.}
    \label{tab:functions}
\end{table}

In order to test the functionality of the library, two well known datasets have been used: the hospital dataset (defined as given in Table~\ref{tab:hospital_example}) and the adult dataset \cite{misc_adult_2}. The simplest one was used for the first proofs of concept due to its simplicity (hospital dataset), while the second one was used to perform multiple tests in a real scenario with a dataset with more than 30K rows (adult dataset). Specifically, with respect to the adult dataset in Table~\ref{tab:adult_extraction} we show an extraction of ten rows and eight columns that have been used as ID, QI and SA during the testing phase:

\begin{table}[ht]
    \centering
    \resizebox{\linewidth}{!}{
    \begin{tabular}{cccccccc}
    \toprule
    \textbf{\textit{age}} & \textbf{\textit{education}}  & \textbf{\textit{marital-status}} & \textbf{\textit{occupation}} & \textbf{\textit{sex}} & \textbf{\textit{native-country}} & \textbf{\textit{race}} & \textbf{\textit{salary-class}}\\
    \midrule
    23 & Some-college & Never-married & Protective-serv & Male & United-States & White & <=50K\\
    60 & HS-grad & Widowed & Other-service & Female & United-States & White & <=50K\\
    60 & Assoc-voc & Widowed & Other-service & Female & United-States & Black & <=50K\\
    48 & Assoc-acdm & Married-civ-spouse & Prof-specialty & Female & United-States & White & >50K\\
    35 & Bachelors & Separated & Craft-repair & Male & United-States & White & <=50K\\
    53 & Some-college & Widowed & Adm-clerical & Female & United-States & Black & <=50K\\
    28 & Some-college & Married-civ-spouse & Protective-serv & Male & United-States & White & <=50K\\
    19 & HS-grad & Never-married & Handlers-cleaners & Male & United-States & White & <=50K\\
    23 & 10th & Never-married & Other-service & Female & United-States & White & <=50K\\
    \bottomrule
    \end{tabular}}
    \caption{Sample extracted from the adult dataset with a selected subset of columns.}
    \label{tab:adult_extraction}
\end{table}

Given the dataset presented in Table~\ref{tab:adult_extraction}, we can consider \textit{race} as identifier in order to suppress it and the \textit{salary-class} as sensitive attribute (since it is the attribute that we want to prevent an attacker from extracting). In addition, the other columns (\textit{age}, \textit{education}, \textit{marital-status}, \textit{occupation}, \textit{sex} and \textit{native-country}) will be the quasi-identifiers. Note that this data extraction does not contain identifying information. 
Therefore, in order to apply the anonymity techniques previously exposed, we will need to introduce the list of QI, ID and the SA, together with the hierarchies to be applied to the quasi-identifiers, the maximum level of records suppression allowed and the anonymity level desired. 
As an illustrative example, let's start by applying \textit{k-anonymity} for $k=10$, \textit{$\ell$-diversity} for $\ell=2$ and \textit{t-closeness} for $t=0.5$ using \textit{anjana}. The steps to follow are detailed in the code given in Code~\ref{code:kanoanymity}, once the hierarchies presented in \cite{sainzpardo2023comparison} have been imported. Regarding creating and defining the hierarchies more details will be given in the \hyperref[sec:methods]{Methods\ref*{sec:methods}} Section.

\begin{code}
\begin{minted}[fontsize=\small]{python}
import pandas as pd
import anjana
from anjana.anonymity import k_anonymity, l_diversity, t_closeness

# Read and process the data
data = pd.read_csv("adult.csv") 
data.columns = data.columns.str.strip()
cols = [
    "workclass",
    "education",
    "marital-status",
    "occupation",
    "sex",
    "native-country",
]
for col in cols:
    data[col] = data[col].str.strip()

# Define the identifiers, quasi-identifiers and the sensitive attribute
quasi_ident = [
    "age",
    "education",
    "marital-status",
    "occupation",
    "sex",
    "native-country",
]
ident = ["race"]
sens_att = "salary-class"

# Select the desired level of k, l and t
k = 10
l = 2
t = 0.5

# Select the suppression limit allowed
supp_level = 50

# Import the hierarchies for each quasi-identifier. Define a dictionary containing them
hierarchies = {
    "age": dict(pd.read_csv("hierarchies/age.csv", header=None)),
    "education": dict(pd.read_csv("hierarchies/education.csv", header=None)),
    "marital-status": dict(pd.read_csv("hierarchies/marital.csv", header=None)),
    "occupation": dict(pd.read_csv("hierarchies/occupation.csv", header=None)),
    "sex": dict(pd.read_csv("hierarchies/sex.csv", header=None)),
    "native-country": dict(pd.read_csv("hierarchies/country.csv", header=None)),
   }

# Apply k-anonymity:
data_kanon = k_anonymity(
    data, ident, quasi_ident, k, supp_level, hierarchies
)

# Apply l-diversity once k-anonymity is implemented:
data_kanon_ldiv = l_diversity(
    data_kanon, ident, quasi_ident, sens_att, k, l, supp_level, hierarchies
)

# Apply t-closeness once k-anonymity and l-diversity are implemented:
data_kanon_ldiv_tclos = t_closeness(
    data_kanon_ldiv, ident, quasi_ident, sens_att, k, t, supp_level, hierarchies
)
\end{minted}
\caption{Example: applying \textit{k-anonymity}, \textit{$\ell$-diversity} and \textit{t-closeness} for the adult dataset ($k=10$, $\ell=2$ and $t=0.5$).}
\label{code:kanoanymity}
\end{code}
\vspace{0.3cm}

The result in this case will be an anonymized dataset verifying \textit{k-anonymity} for $k=10$, $\ell=2$ for \textit{$\ell$-diversity} and $t=0.5$ for \textit{t-closeness}, which can be stored as a \textit{csv} file or other format for later use. Note that this dataset contains 32561 rows and the application of this technique using \textit{anjana} provides the anonymized dataset in less than one second (regarding the computational resources available during the test phase\footnote{For testing and developing the library a machine provided with 12th Gen Intel(R) Core(TM) i5-1 and with 16 GB RAM running under Ubuntu 22.04.4 LTS was used.}). Specifically, applying \codehighlight{pycanon}\cite{pycanon_paper} we get the following values once anonymized with the three techniques: $k=72$, $\ell=2$ and $t=0.47370$.

During the functionality test phase and examples, all the techniques of the library have been applied to this dataset with the values of Table~\ref{tab:anonymity_pycanon}. The actual value calculated using the \codehighlight{pycanon} library is shown in that table. It should be noted that for \textit{entropy $\ell$-diversity} and \textit{recursive (c,$\ell$)-diversity} it has not been possible to anonymize it given the hierarchies introduced, so they are not shown. In addition, the percentage of records eliminated in each case is shown.

\begin{table}[ht]
    \centering
    \begin{tabular}{crrr}
    \toprule
    \textbf{\textit{Technique}} & \textbf{\textit{Anonymity values applied}} & \textbf{\textit{Anonymity values calculated}} & \textbf{\textit{Suppressed records ($\%$)}} \\
    \midrule
    \textit{k-anonymity} & $k=10$, \textit{supp\_level}$=50\%$ & $k=10$ & 43.71 $\%$\\
    \textit{($\alpha$,k)-anonymity} & $k=10$, $\alpha=0.8$, \textit{supp\_level}$=100\%$ & $k=10$, $\alpha=0.8$ & 87.99 $\%$ \\
    \textit{$\ell$-diversity} & $k=10$, $\ell=2$, \textit{supp\_level}$=50\%$ & $k=72$, $\ell=2$ & 43.71 $\%$\\
    \textit{t-closeness} & $k=10$, $t=0.5$, \textit{supp\_level}$=50\%$ & $k=72$, $t=0.4737$ & 43.71 $\%$\\
    \textit{$\delta$-disclosure privacy} & $k=10$, $\delta=3$, \textit{supp\_level}$=50\%$  & $k=392$, $\delta=2.159$ & 43.71 $\%$\\
    \textit{Basic $\beta$-likeness} & $k=10$, $\beta=0.5$, \textit{supp\_level}$=100\%$ & $k=2098$, $\beta=0.4178$ & 72.74 $\%$\\
    \textit{Enhanced $\beta$-likeness} & $k=10$, $\beta=0.5$, \textit{supp\_level}$=100\%$ & $k=2098$, $\beta=0.4178$ & 72.74 $\%$\\
    \bottomrule
    \end{tabular}
    \caption{Anonymity techniques applied to the adult dataset: values for the privacy suppression levels applied using \textit{anjana}, real anonymity values calculated using \textit{pyCANON} and percentage of suppressed records.}
    \label{tab:anonymity_pycanon}
\end{table}

\subsection*{Software development}

\paragraph{Dependencies management and installation}
To manage the dependencies of the library and its installation, \codehighlight{poetry}\cite{poetry} has been used. Specifically, a \textit{pyproject.toml} file has been created which, in addition to these dependencies, contains the metadata associated with the library (version, authors, license, classification, etc). The dependencies are distinguished between those that must be installed together with the library for its correct operation (described in  \codehighlight{[tool.poetry.dependencies]}), and the dependencies necessary to run the tests, create the documentation, check the code style, etc.

\paragraph{CI/CD and publish pipelines}
In the current version, the development of the library is done on GitHub and GitHub actions have been used for creating a CI/CD pipeline and a workflow for publishing the current release in PyPI. The CI/CD pipeline (named \textit{cicd.yml}) is executed with the following stages: lint, test and build. The lint stage is responsible for checking the code style and syntax errors. In the test stage we check the correct execution in Python 3.9 3.10 3.11 and 3.12. Finally, the build stage checks that the package is built correctly and the dependencies are properly installed.

\paragraph{Unit and functional tests}
Regarding the validation of the correct functionality of the library, two types of tests have been carried out: unit and functional tests. In the first ones, we have tested the boundary cases, the correct management of exceptions, the correct typing of the function inputs, and the extreme cases. In the functionality tests, we have verified that the implemented anonymity functions work correctly, returning the expected results based on the desired level of anonymity applied. For the latter, the \codehighlight{pycanon} library has been used as auxiliary library, since it already has its own tests that ensure the proper performance of the same. To implement these tests \codehighlight{pytest}\cite{pytest} (version >= 7.1.2 and < 9.0.0) has been used. These tests are located in the test folder of the library's repository. 

Additionally, we are interested in determining the level of code coverage obtained with these tests, for which \codehighlight{codecov}\cite{codecov} has been used. In this sense, a workflow has been created in GitHub using a YML pipeline that creates a Python 3.10 environment, installs \codehighlight{pytest}, \codehighlight{poetry}, the dependencies and uses the \codehighlight{coverage run} command to run the tests and obtain the code coverage. Subsequently, using a codecov token, the code coverage level is published and can be analyzed within the \href{https://app.codecov.io/gh/IFCA-Advanced-Computing/anjana}{interactive interface}, along with the scan that shows the lines that are fully covered, partially covered and uncovered. For version 1.0.0, the code coverage is greater than 92$\%$.

\paragraph{Release automation} The \codehighlight{release-please}\cite{release_please} library has been used to create automated releases. A workflow has been defined using GitHub actions to create the releases when there is a push on the main branch. 
Finally, in case a new tag has been included, it is automatically published in PyPI, so that the installation can simply be carried out running the  \codehighlight{pip install anjana} command.

\paragraph{Documentation} To create the library documentation, \codehighlight{sphinx}\cite{brandl2021sphinx} has been used, in order to generate the documentation of the implemented functions from the docstrings. For this purpose, the docstrings have been written using reStructuredText. Moreover, additional pages have been included to the documentation such as an installation and getting started guide, information about hierarchy management and how to manage multiple sensitive attributes, among others. The documentation has been published in readthedocs in an automated way and it is available in: \url{https://anjana.readthedocs.io/en/latest/}.

\paragraph{Requirements} In addition to the requirements for the tests, lint and build phases, the requirements for the proper operation of the library are the following: \codehighlight{Python}\cite{10.5555/1593511} (version >= 3.9), \codehighlight{numpy}\cite{harris2020array} (version 2.0.1), \codehighlight{pandas}\cite{mckinney-proc-scipy-2010} (version 2.2.2), \codehighlight{pycanon}\cite{pycanon_paper} (version 1.0.1.post2), \codehighlight{typing\_extensions}\cite{typing_extensions} (version 4.12.2), \codehighlight{beartype}\cite{beartype} (version 0.18.5) and \codehighlight{docutils}\cite{docutils} (version 0.21.2). Note that \codehighlight{dependabot} has been used for keeping the dependencies up to date, so the versions of the dependencies may change (see the workflow implemented in \href{https://github.com/IFCA-Advanced-Computing/anjana/blob/main/.github/dependabot.yml}{dependabot.yml}).

\section*{Discussion}\label{sec:discussion}
\subsection*{Multiple sensitive attributes}

In version 1.0.0, \textit{anjana} allows the incorporation of a single sensitive attribute (SA) for the anonymization process with the different techniques implemented. If there are more than one SA, two approaches can be followed as stated in \cite{pycanon_paper}: \textit{harmonization of the quasi-identifiers} or \textit{quasi-identifiers update}. In the first one we should apply each method to each sensitive attribute recursively, i.e., we perform the anonymization for the first SA and then anonymize the resulting dataset with respect to the next one, and so on until all of them are covered. In the second case (\textit{quasi-identifiers update}) it is considered that some of the sensitive attributes can act as a quasi-identifier for the rest of the sensitive attributes. In this case, we shall anonymize the data for the first SA adding to the list of quasi-identifiers the remaining sensitive attributes that may act as QIs in the case of the current SA. Once anonymized for that SA, the process is repeated for the remaining ones updating in each case the list of QIs according to the data and attributes requirements.

\subsection*{Creating the hierarchies}
Regarding the hierarchies used for generalizing the quasi-identifiers, all the anonymity functions available in \textit{anjana} receive a dictionary with the hierarchies to be applied to the QIs. In particular, this dictionary has as key the names of the columns that are QIs to which a hierarchy is to be applied (it may happen that a user does not want to generalize some QIs and therefore no hierarchy is to be applied to them, just do not include them in this dictionary). The value for each key (QI) is formed by a dictionary in such a way that the value 0 has as value the raw column (as it is in the original dataset), the value 1 corresponds to the first level of transformation to be applied, in relation to the values of the original column, and so on with as many keys as levels of hierarchies have been established.

For a better understanding, let's look at the following example. Suppose that we have the following simulated dataset given in Table~\ref{tab:hospital_example} (corresponding with the \textit{hospital\_extended.csv} dataset used for testing purposes) with \textit{age}, \textit{gender} and \textit{city} as quasi-identifiers, \textit{name} as identifier, \textit{disease} as SA and \textit{religion} as insensitive attribute. Regarding the QIs, we want to apply the following hierarchies: interval of 5 years (first level) and 10 years (second level) for the \textit{age}. Suppression as first (and only) level for both \textit{gender} and \textit{city}.

\begin{table}[ht]
    \centering
    \begin{tabular}{cccccc}
    \toprule
        \textbf{\textit{name}} & \textbf{\textit{age}}  & \textbf{\textit{gender}}  & \textbf{\textit{city}} & \textbf{\textit{religion}} & \textbf{\textit{disease}}\\
        \midrule
        Ramsha & 29 & Female & Tamil Nadu & Hindu & Cancer \\
        Gabu & 24 & Male & Tamil Nadu & Hindu & Cancer \\
        Sabu & 23 & Male & Tamil Nadu & Hindu & Cancer \\
        Jonas & 22 & Male & Tamil Nadu & Hindu & Cancer \\
        Yadu & 24 & Female & Kerala & Hindu & Viral infection \\
        Salima & 28 & Female & Tamil Nadu & Muslim & TB \\
        Sunny & 27 & Male & Karnataka & Parsi & No illness \\
        Joan & 24 & Female & Kerala & Christian & Heart-related \\
        Bahuksana & 23 & Male & Karnataka & Buddhist & TB \\
        Rambha & 19 & Male & Kerala & Hindu & Cancer \\
        Kishor & 29 & Male & Karnataka & Hindu & Heart-related \\
        Johnson & 17 & Male & Kerala & Christian & Heart-related \\
        John & 19 & Male & Kerala & Christian & Viral infection \\
    \bottomrule
    \end{tabular}
    \caption{Example: data from the \textit{hospital\_extended.csv} file}
    \label{tab:hospital_example}
\end{table}

Then, in order to create the hierarchies we can define the following dictionary for \textit{age}, \textit{gender} and \textit{city}, using the auxiliary function \textit{generate\_intervals()} available in the utils subpackage form the library to generate interval-based hierarchies. This function receives as input the original data to be generalized, the lower and upper bounds for the interval set and the spacing between the values of the intervals (step). Note that following the requirement previously mentioned, for the quasi-identifiers \textit{gender} and \textit{city} only suppression is applied, while for \textit{age} intervals of five and ten years are used as generalization levels 1 and 2 respectively, as stated in Code~\ref{code:hierarchies}.

\newpage
\begin{code}
\begin{minted}[fontsize=\small]{python}
import numpy as np 
from anajana.anonymity import utils

hierarchies = {
    "age": {
        0: data["age"].values,
        1: utils.generate_intervals(data["age"].values, 0, 100, 5),
        2: utils.generate_intervals(data["age"].values, 0, 100, 10),
    },
    "gender": {
        0: data["gender"].values,
        1: np.array(["*"] * len(data["gender"].values)) # Suppression
    },
    "city": {
        0: data["city"].values,
        1: np.array(["*"] * len(data["city"].values)) # Suppression
    } 
}
\end{minted}
\caption{Example: creating hierarchies for the quasi-idenfiers of a database.}
\label{code:hierarchies}
\end{code}
\vspace{0.3cm}

In view of the hierarchies created in Code~\ref{code:hierarchies} for the three quasi-identifiers, if we anonymize Table~\ref{tab:hospital_example} with \textit{name} as identifier, \textit{age}, \textit{gender} and \textit{city} as quasi-identifiers, and \textit{disease} as SA, using a suppression level of 0$\%$ and \textit{k-anonymity} with $k=2$, the resulting dataset is shown in Table~\ref{tab:hospital_example_k2}. In addition, Table~\ref{tab:hospital_example_l2} shows the resulting dataset when in addition to \textit{k-anonymity} with $k=2$, \textit{$\ell$-diversity} is also applied with $\ell=2$. For the first one, only the second level of generalization has been applied to \textit{age} (intervals of 10 years), and for the second one it has also been applied suppression for the QI \textit{city} (first level of generalization created).

\begin{table}[ht]
    \centering
    \begin{tabular}{cccccc}
    \toprule
        \textbf{\textit{name}} & \textbf{\textit{age}}  & \textbf{\textit{gender}}  & \textbf{\textit{city}} & \textbf{\textit{religion}} & \textbf{\textit{disease}}\\
        \midrule
        * & [20, 30) & Female & Tamil Nadu & Hindu & Cancer \\
        * & [20, 30) & Male & Tamil Nadu & Hindu & Cancer \\
        * & [20, 30) & Male & Tamil Nadu & Hindu & Cancer \\
        * & [20, 30) & Male & Tamil Nadu & Hindu & Cancer \\
        * & [20, 30) & Female & Kerala & Hindu & Viral infection \\
        * & [20, 30) & Female & Tamil Nadu & Muslim & TB \\
        * & [20, 30) & Male & Karnataka & Parsi & No illness \\
        * & [20, 30) & Female & Kerala & Christian & Heart-related \\
        * & [20, 30) & Male & Karnataka & Buddhist & TB \\
        * & [10, 20) & Male & Kerala & Hindu & Cancer \\
        * & [20, 30) & Male & Karnataka & Hindu & Heart-related \\
        * & [10, 20) & Male & Kerala & Christian & Heart-related \\
        * & [10, 20) & Male & Kerala & Christian & Viral infection \\
    \bottomrule
    \end{tabular}
    \caption{Example: \textit{hospital\_extended.csv} anonymizing with \textit{k-anonymity} with $k=2$.}
    \label{tab:hospital_example_k2}
\end{table}

\begin{table}[ht]
    \centering
    \begin{tabular}{cccccc}
    \toprule
        \textbf{\textit{name}} & \textbf{\textit{age}}  & \textbf{\textit{gender}}  & \textbf{\textit{city}} & \textbf{\textit{religion}} & \textbf{\textit{disease}}\\
        \midrule
         * & [20, 30) & Female & * & Hindu & Cancer \\
         * & [20, 30) & Male & * & Hindu & Cancer \\
         * & [20, 30) & Male & * & Hindu & Cancer \\
         * & [20, 30) & Male & * & Hindu & Cancer \\
         * & [20, 30) & Female & * & Hindu & Viral infection \\
         * & [20, 30) & Female & * & Muslim & TB \\
         * & [20, 30) & Male & * & Parsi & No illness \\
         * & [20, 30) & Female & * & Christian & Heart-related \\
         * & [20, 30) & Male & * & Buddhist & TB \\
         * & [10, 20) & Male & * & Hindu & Cancer \\
         * & [20, 30) & Male & * & Hindu & Heart-related \\
         * & [10, 20) & Male & * & Christian & Heart-related \\
         * & [10, 20) & Male & * & Christian & Viral infection \\
    \bottomrule
    \end{tabular}
    \caption{Example: \textit{hospital\_extended.csv} anonymizing with \textit{k-anonymity} with $k=2$ and \textit{$\ell$-diversity} with $\ell=2$. Note that this table verifies \textit{k-anonymity} for $k=3$.}
    \label{tab:hospital_example_l2}
\end{table}

\subsection*{Anonymity transformation applied}
Regarding the transformation applied for anonymizing the data based on the hierarchy level applied to each quasi-identifier, in some cases it is important to extract it in order to transfer statistics on the processing performed on the data. Usually, this transformation will be denoted with a list of the same length as the number of quasi-identifiers. When performing the anonymization process, the quasi-identifiers are entered in a certain order, which will be the same as the order in which they are represented in the list with the transformation.

To obtain this transformation, the \textit{get\_transformation()} function from the \textit{utils} sub-module can be used introducing the anonymizing dataset, the set of quasi-identifiers and the dictionary containing the hierarchies. This can be of particular interest in many cases, such as to transmit statistics about the anonymization process performed without sharing the data. Additionally, in an architecture where different data owners are anonymizing different datasets following the same hierarchies, if once the data is anonymized they will want to train a global model, either in a distributed way (for example under a federated learning architecture) or in a centralized way, it will be necessary to harmonize the transformations carried out. Thus, it will be of interest that all data owners have applied the same transformations to the quasi-identifiers in order to train the model (and, if necessary, aggregate it) in a consistent way.

Specifically, the function \textit{get\_transformation()} returns a list so that the value in the position $i$ corresponds to the hierarchy level applied for the quasi-identifier entered in position $i$ in the list of quasi-identifiers entered during the anonymization process, $\forall i \{1, \hdots, n_{QI}\}$, with $n_{QI}$ the number of quasi-identifiers introduced. For example, a transformation $[1,0,2,3]$ represents that the first level of hierarchy has been applied to the first QI, that no transformation has been applied to the second, and that the second and third levels of hierarchies have been applied respectively to the third and fourth quasi-identifiers. 

In the example defined above, note that the transformation applied to Table~\ref{tab:hospital_example} for obtaining Table~\ref{tab:hospital_example_k2} is [2,0,0], while for obtaining Table~\ref{tab:hospital_example_l2} the transformation is [2,0,1].

\subsection*{Anonymization in data science workflows} 
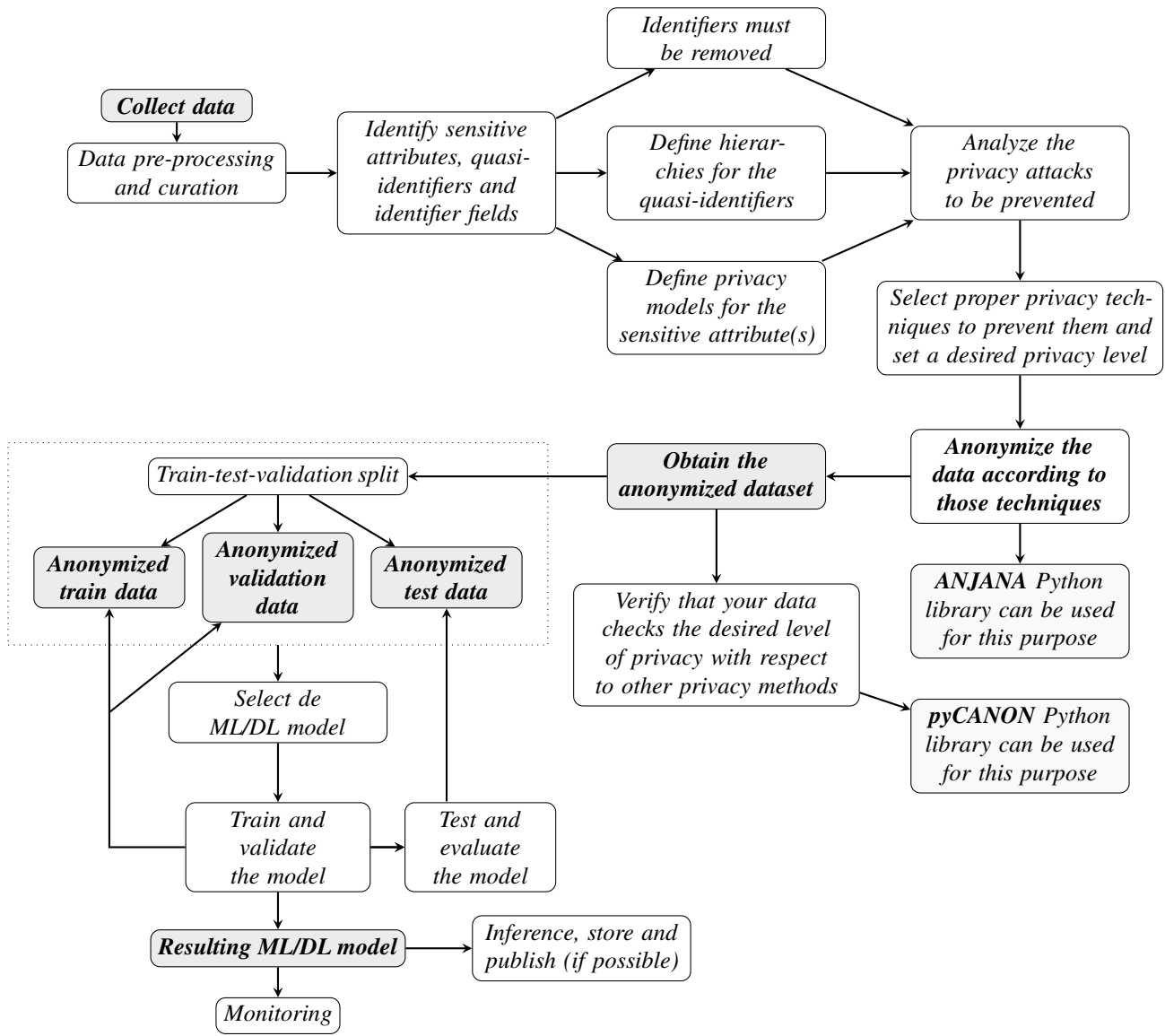
\begin{figure}[ht!]
\begin{tikzpicture}
    \node[draw, rectangle, rounded corners, text width=2cm, text centered, fill=gray!15] (A) at (0,0) {\textit{\textbf{Collect data}}};
    \node[draw,rectangle,rounded corners, text width=3cm, text centered] (B) at (0,-1) {\textit{Data pre-processing and curation}};
    \node[draw,rectangle,rounded corners, text width=3cm, text centered] (C) at (4,-1) {\textit{Identify sensitive attributes, quasi-identifiers and identifier fields}};
    \node[draw,rectangle,rounded corners, text width=3cm, text centered] (D) at (8,1) {\textit{Identifiers must be removed}};
    \node[draw,rectangle,rounded corners, text width=3cm, text centered] (E) at (8,-1) {\textit{Define hierarchies for the quasi-identifiers}};
    \node[draw,rectangle,rounded corners, text width=3cm, text centered] (F) at (8,-3) {\textit{Define privacy models for the sensitive attribute(s)}};
    \node[draw,rectangle,rounded corners, text width=3cm, text centered] (G) at (12.5,-1) {\textit{Analyze the privacy attacks to be prevented}};
    \node[draw,rectangle,rounded corners, text width=4cm, text centered] (H) at (12.5,-3.3) {\textit{Select proper privacy techniques to prevent them and set a desired privacy level}};
    \node[draw,rectangle,rounded corners, text width=3cm, text centered] (I) at (12.5,-5.5) {\textit{\textbf{Anonymize the data according to those techniques}}};
    \node[draw,rectangle,rounded corners, text width=3cm, text centered, fill=gray!5] (J) at (12.5,-7.5) {\textit{\textbf{ANJANA} Python library can be used for this purpose}};
    \node[draw,rectangle,rounded corners, text width=3cm, text centered, fill=gray!15] (K) at (8,-5.5) {\textit{\textbf{Obtain the anonymized dataset}}};
    \node[draw,rectangle,rounded corners, text width=4cm, text centered] (L) at (8,-8) {\textit{Verify that your data checks the desired level of privacy with respect to other privacy methods}};
    \node[draw,rectangle,rounded corners, text width=3cm, text centered, fill=gray!5] (M) at (12.5,-9.5) {\textit{\textbf{pyCANON} Python library can be used for this purpose}};
    \node[draw,rectangle,rounded corners, text centered] (N) at (1.5,-5.5) {\textit{Train-test-validation split}};
    \node[draw,rectangle,rounded corners, text width=2cm, text centered, fill=gray!15] (O) at (-1,-7) {\textit{\textbf{Anonymized train data}}};
    \node[draw,rectangle,rounded corners, text width=2cm, text centered, fill=gray!15] (P) at (1.5,-7) {\textit{\textbf{Anonymized validation data}}};
    \node[draw,rectangle,rounded corners, text width=2cm, text centered, fill=gray!15] (Q) at (4,-7) {\textit{\textbf{Anonymized test data}}};
    \node[draw,rectangle,rounded corners, text width=3cm, text centered] (R) at (1.5,-9) {\textit{Select de ML/DL model}};
    \node[draw,rectangle,rounded corners, text width=2.5cm, text centered] (S) at (1.5,-11) {\textit{Train and validate the model}};
    \node[draw,rectangle,rounded corners, text centered, text width=2cm] (T) at (4.5,-11) {\textit{Test and evaluate the model}};
    \node[draw,rectangle,rounded corners, text centered, fill=gray!15] (U) at (1.5,-12.5) {\textit{\textbf{Resulting ML/DL model}}};
    \node[draw,rectangle,rounded corners, text centered] (V) at (1.5,-13.5) {\textit{Monitoring}};
    \node[draw,rectangle,rounded corners, text width=3cm, text centered] (W) at (6,-12.5) {\textit{Inference, store and publish (if possible)}};  

    \draw[dotted] (-2.5,-8) rectangle (5.5,-5);

    \draw[thick,->,>=stealth] (A) -- (B);
    \draw[thick,->,>=stealth] (B) -- (C);
    \draw[thick,->,>=stealth] (C) -- (D);
    \draw[thick,->,>=stealth] (C) -- (E);
    \draw[thick,->,>=stealth] (C) -- (F);
    \draw[thick,->,>=stealth] (D) -- (G);
    \draw[thick,->,>=stealth] (E) -- (G);
    \draw[thick,->,>=stealth] (F) -- (G);
    \draw[thick,->,>=stealth] (G) -- (H);
    \draw[thick,->,>=stealth] (H) -- (I);
    \draw[thick,->,>=stealth] (I) -- (J);
    \draw[thick,->,>=stealth] (I) -- (K);
    \draw[thick,->,>=stealth] (K) -- (L);
    \draw[thick,->,>=stealth] (L) -- (M);
    \draw[thick,->,>=stealth] (K) -- (N);
    \draw[thick,->,>=stealth] (N) -- (O);
    \draw[thick,->,>=stealth] (N) -- (P);
    \draw[thick,->,>=stealth] (N) -- (Q);
    \draw[thick,->,>=stealth] (1.5,-8) -- (R);
    \draw[thick,->,>=stealth] (R) -- (S);
    \draw[thick,->,>=stealth] (S) -- ++(-1.4,0) -| (O);
    \draw[thick,->,>=stealth] ++(-1,-9) -- (P);
    \draw[thick,->,>=stealth] (S) -- (T);
    \draw[thick,->,>=stealth] (4,-10.32) -- (Q);
    \draw[thick,->,>=stealth] (S) -- (T);
    \draw[thick,->,>=stealth] (S) -- (U);
    \draw[thick,->,>=stealth] (U) -- (V);
    \draw[thick,->,>=stealth] (U) -- (W);
    \end{tikzpicture}
\caption{Classic workflow of a data-based AI project including the training/validation and testing phase of ML/DL models and the data anonymization process.}
\label{fig:diagram_ml_process}
\end{figure}

Applying the previously exposed anonymity techniques appropriately with special attention to the privacy attacks that are to be prevented is a fundamental step in a data science process. In particular, the diagram in Figure~\ref{fig:diagram_ml_process} details the flow of a machine or deep learning project, including the correct management of the privacy of the data involved.

From the diagram in Figure~\ref{fig:diagram_ml_process}, several factors must be taken into account: first, the fact that the initial dataset verifies a certain privacy constraint does not mean that once it has been split into train/val/test, it will continue to be verified. Therefore, it may be interesting to perform the train test split before anonymizing. In the same sense, if the train test split is performed before anonymizing, it should be considered that in order to train, validate and evaluate the ML/DL model, it is convenient that the same transformations have been applied at the time of anonymizing the data. In other words, if the model has been trained with data transformed by applying certain hierarchies, the validation and test data must have been anonymized with the same levels of hierarchies. It is important to note that these hierarchies may not allow to reach the desired level of anonymity, so it is necessary to be quite conservative in this part. Thus, depending on the objective and scope of the model to be developed, it will be interesting to follow the scheme shown in Figure~\ref{fig:diagram_ml_process} or to invert it so that the anonymization is performed just before the train/validation/test split, harmonizing in each case the transformations applied in terms of the hierarchies on the quasi-identifiers (taking the strictest one in order to guarantee the required privacy level for the three data splits).

\subsection*{Future work}

This paper presents the implementation of an open source Python library that allows to anonymize tabular data using the most commonly used techniques in an intuitive and user-friendly way. This tool aims to address the need to provide researchers and general users with a comprehensive tool that allows them to locally anonymize their sensitive data using Python, so that it can be easily integrated into the workflow of an ML/DL project. For this purpose, nine anonymization techniques can be applied on tabular data by introducing the data, identifiers, quasi-identifiers, sensitive attribute, maximum level of records' suppression allowed and privacy level required according with the method to be applied.

Regarding future lines of work, it is of particular interest to include additional anonymity tools together with metrics to quantify the loss of information during the anonymization process, as well as to analyze the risk of attribute disclosure. Also, in order to make the tool even easier and more attractive, the creation of an application that allows users to carry out the anonymization process through a local interface or dashboard (so that the data does not have to be uploaded to any web or external service) will also be explored. Finally, identifying ways to quantify resistance to attacks based on auxiliary information is an interesting line of work, as well as incorporating additional techniques, including pseudonymization of identifiers.

\section*{Data Availability}
No new data have been generated for this work. The data used to test the library and the corresponding anonymized data generated according to some techniques tested can be found in the \href{https://github.com/IFCA-Advanced-Computing/anjana/tree/develop/examples}{examples folder of the library's repository}.

\section*{Code availability}
Relevant links regarding the availability of the source code, documentation and installation of the software are listed below.
\begin{itemize}
    \item The code for the \textit{anjana} library is openly available in GitHub:
    \href{https://github.com/IFCA-Advanced-Computing/anjana}{https://github.com/IFCA-Advanced-Computing/anjana}.
    \item The documentation can be found in ReadTheDocs: \href{https://anjana.readthedocs.org}{https://anjana.readthedocs.org}.
    \item A Zenodo DOI has been created to ensure code reproducibility and availability. This DOI stands for all versions, and always resolves to the latest one: \href{https://zenodo.org/doi/10.5281/zenodo.11184467}{https://zenodo.org/doi/10.5281/zenodo.11184467}.
\end{itemize}

\bibliography{main_arxiv}

\begin{thebibliography}{10}
\urlstyle{rm}
\expandafter\ifx\csname url\endcsname\relax
  \def\url#1{\texttt{#1}}\fi
\expandafter\ifx\csname urlprefix\endcsname\relax\def\urlprefix{URL }\fi
\expandafter\ifx\csname doiprefix\endcsname\relax\def\doiprefix{DOI: }\fi
\providecommand{\bibinfo}[2]{#2}
\providecommand{\eprint}[2][]{\url{#2}}

\bibitem{10.1145/3457607}
\bibinfo{author}{Mehrabi, N.}, \bibinfo{author}{Morstatter, F.}, \bibinfo{author}{Saxena, N.}, \bibinfo{author}{Lerman, K.} \& \bibinfo{author}{Galstyan, A.}
\newblock \bibinfo{journal}{\bibinfo{title}{A survey on bias and fairness in machine learning}}.
\newblock {\emph{\JournalTitle{ACM Comput. Surv.}}} \textbf{\bibinfo{volume}{54}}, \url{10.1145/3457607} (\bibinfo{year}{2021}).

\bibitem{10.1145/3631326}
\bibinfo{author}{Hort, M.}, \bibinfo{author}{Chen, Z.}, \bibinfo{author}{Zhang, J.~M.}, \bibinfo{author}{Harman, M.} \& \bibinfo{author}{Sarro, F.}
\newblock \bibinfo{journal}{\bibinfo{title}{Bias mitigation for machine learning classifiers: A comprehensive survey}}.
\newblock {\emph{\JournalTitle{ACM J. Responsib. Comput.}}} \url{10.1145/3631326} (\bibinfo{year}{2023}).

\bibitem{GDPR2016a}
\bibinfo{author}{{European Commission}}.
\newblock \bibinfo{title}{Regulation of the european parliament and of the council of 27 april 2016 on the protection of natural persons with regard to the processing of personal data and on the free movement of such data}.
\newblock \bibinfo{howpublished}{\url{https://data.europa.eu/eli/reg/2016/679/oj}}.
\newblock \bibinfo{note}{[Accessed 16-05-2024]}.

\bibitem{AIAct}
\bibinfo{author}{{European Commission}}.
\newblock \bibinfo{title}{Regulation of the european parliament and of the council laying down harmonised rules on artificial intelligence (artificial intelligence act) and emending certain union legislative acts}.
\newblock \bibinfo{howpublished}{\url{https://eur-lex.europa.eu/legal-content/EN/TXT/?uri=CELEX:52021PC0206}}.
\newblock \bibinfo{note}{[Accessed 16-05-2024]}.

\bibitem{PrivacyAct}
\bibinfo{author}{{Office of Privacy and Civil Liberties. U.S. Department of Justice}}.
\newblock \bibinfo{title}{Privacy act of 1974}.
\newblock \bibinfo{howpublished}{\url{https://www.justice.gov/opcl/privacy-act-1974}}.
\newblock \bibinfo{note}{[Accessed 16-05-2024]}.

\bibitem{hipaa}
\bibinfo{author}{{Centers for Medicare \& Medicaid Services}}.
\newblock \bibinfo{title}{{The Health Insurance Portability and Accountability Act of 1996 (HIPAA)}}.
\newblock \bibinfo{howpublished}{Online at http://www.cms.hhs.gov/hipaa/} (\bibinfo{year}{1996}).

\bibitem{sweeney2002k}
\bibinfo{author}{Sweeney, L.}
\newblock \bibinfo{journal}{\bibinfo{title}{k-anonymity: A model for protecting privacy}}.
\newblock {\emph{\JournalTitle{International journal of uncertainty, fuzziness and knowledge-based systems}}} \textbf{\bibinfo{volume}{10}}, \bibinfo{pages}{557--570} (\bibinfo{year}{2002}).

\bibitem{wong2006alpha}
\bibinfo{author}{Wong, R. C.-W.}, \bibinfo{author}{Li, J.}, \bibinfo{author}{Fu, A. W.-C.} \& \bibinfo{author}{Wang, K.}
\newblock \bibinfo{title}{($\alpha$, k)-anonymity: an enhanced k-anonymity model for privacy preserving data publishing}.
\newblock In \emph{\bibinfo{booktitle}{Proceedings of the 12th ACM SIGKDD international conference on Knowledge discovery and data mining}}, \bibinfo{pages}{754--759} (\bibinfo{year}{2006}).

\bibitem{machanavajjhala2007diversity}
\bibinfo{author}{Machanavajjhala, A.}, \bibinfo{author}{Kifer, D.}, \bibinfo{author}{Gehrke, J.} \& \bibinfo{author}{Venkitasubramaniam, M.}
\newblock \bibinfo{journal}{\bibinfo{title}{l-diversity: Privacy beyond k-anonymity}}.
\newblock {\emph{\JournalTitle{Acm transactions on knowledge discovery from data (tkdd)}}} \textbf{\bibinfo{volume}{1}}, \bibinfo{pages}{3--es} (\bibinfo{year}{2007}).

\bibitem{li2006t}
\bibinfo{author}{Li, N.}, \bibinfo{author}{Li, T.} \& \bibinfo{author}{Venkatasubramanian, S.}
\newblock \bibinfo{title}{t-closeness: Privacy beyond k-anonymity and l-diversity}.
\newblock In \emph{\bibinfo{booktitle}{2007 IEEE 23rd international conference on data engineering}}, \bibinfo{pages}{106--115} (\bibinfo{organization}{IEEE}, \bibinfo{year}{2006}).

\bibitem{10.1145/1401890.1401904}
\bibinfo{author}{Brickell, J.} \& \bibinfo{author}{Shmatikov, V.}
\newblock \bibinfo{title}{The cost of privacy: destruction of data-mining utility in anonymized data publishing}.
\newblock In \emph{\bibinfo{booktitle}{Proceedings of the 14th ACM SIGKDD International Conference on Knowledge Discovery and Data Mining}}, KDD '08, \bibinfo{pages}{70–78}, \url{10.1145/1401890.1401904} (\bibinfo{publisher}{Association for Computing Machinery}, \bibinfo{address}{New York, NY, USA}, \bibinfo{year}{2008}).

\bibitem{cao2012publishing}
\bibinfo{author}{Cao, J.} \& \bibinfo{author}{Karras, P.}
\newblock \bibinfo{journal}{\bibinfo{title}{Publishing microdata with a robust privacy guarantee}}.
\newblock {\emph{\JournalTitle{arXiv preprint arXiv:1208.0220}}}  (\bibinfo{year}{2012}).

\bibitem{10.1007/978-3-319-45381-1_5}
\bibinfo{author}{Domingo-Ferrer, J.} \& \bibinfo{author}{Soria-Comas, J.}
\newblock \bibinfo{title}{Anonymization in the time of big data}.
\newblock In \bibinfo{editor}{Domingo-Ferrer, J.} \& \bibinfo{editor}{Peji{\'{c}}-Bach, M.} (eds.) \emph{\bibinfo{booktitle}{Privacy in Statistical Databases}}, \bibinfo{pages}{57--68} (\bibinfo{publisher}{Springer International Publishing}, \bibinfo{address}{Cham}, \bibinfo{year}{2016}).

\bibitem{sklearn_api}
\bibinfo{author}{Buitinck, L.} \emph{et~al.}
\newblock \bibinfo{title}{{API} design for machine learning software: experiences from the scikit-learn project}.
\newblock In \emph{\bibinfo{booktitle}{ECML PKDD Workshop: Languages for Data Mining and Machine Learning}}, \bibinfo{pages}{108--122} (\bibinfo{year}{2013}).

\bibitem{chollet2015keras}
\bibinfo{author}{Chollet, F.} \emph{et~al.}
\newblock \bibinfo{title}{Keras}.
\newblock \bibinfo{howpublished}{\url{https://keras.io}} (\bibinfo{year}{2015}).

\bibitem{tensorflow2015-whitepaper}
\bibinfo{author}{Abadi, M.} \emph{et~al.}
\newblock \bibinfo{title}{{TensorFlow}: Large-scale machine learning on heterogeneous systems} (\bibinfo{year}{2015}).
\newblock \bibinfo{note}{Software available from tensorflow.org}.

\bibitem{NEURIPS2019_9015}
\bibinfo{author}{Paszke, A.} \emph{et~al.}
\newblock \bibinfo{title}{Pytorch: An imperative style, high-performance deep learning library}.
\newblock In \emph{\bibinfo{booktitle}{Advances in Neural Information Processing Systems 32}}, \bibinfo{pages}{8024--8035} (\bibinfo{publisher}{Curran Associates, Inc.}, \bibinfo{year}{2019}).

\bibitem{ayala2014systematic}
\bibinfo{author}{Ayala-Rivera, V.}, \bibinfo{author}{McDonagh, P.}, \bibinfo{author}{Cerqueus, T.}, \bibinfo{author}{Murphy, L.} \emph{et~al.}
\newblock \bibinfo{journal}{\bibinfo{title}{A systematic comparison and evaluation of k-anonymization algorithms for practitioners.}}
\newblock {\emph{\JournalTitle{Trans. Data Priv.}}} \textbf{\bibinfo{volume}{7}}, \bibinfo{pages}{337--370} (\bibinfo{year}{2014}).

\bibitem{yuvaraj2022privacy}
\bibinfo{author}{Yuvaraj, N.}, \bibinfo{author}{Praghash, K.} \& \bibinfo{author}{Karthikeyan, T.}
\newblock \bibinfo{journal}{\bibinfo{title}{Privacy preservation of the user data and properly balancing between privacy and utility}}.
\newblock {\emph{\JournalTitle{International Journal of Business Intelligence and Data Mining}}} \textbf{\bibinfo{volume}{20}}, \bibinfo{pages}{394--411} (\bibinfo{year}{2022}).

\bibitem{anonypy}
\bibinfo{author}{{AnonyPy developers}}.
\newblock \bibinfo{title}{Github repository: anonypy}.
\newblock \bibinfo{howpublished}{\url{https://github.com/glassonion1/anonypy}} (\bibinfo{year}{2024}).
\newblock \bibinfo{note}{[Accessed 08-05-2024]}.

\bibitem{anonym}
\bibinfo{author}{{Anonym developers}}.
\newblock \bibinfo{title}{Github repository: anonym}.
\newblock \bibinfo{howpublished}{\url{https://gitlab.com/datainnovatielab/public/anonym}} (\bibinfo{year}{2024}).
\newblock \bibinfo{note}{[Accessed 08-05-2024]}.

\bibitem{dicognito}
\bibinfo{author}{{Dicognito developers}}.
\newblock \bibinfo{title}{Github repository: dicognito}.
\newblock \bibinfo{howpublished}{\url{https://github.com/blairconrad/dicognito}} (\bibinfo{year}{2024}).
\newblock \bibinfo{note}{[Accessed 08-05-2024]}.

\bibitem{privapy}
\bibinfo{author}{{Privapy developers}}.
\newblock \bibinfo{title}{Github repository: privapy}.
\newblock \bibinfo{howpublished}{\url{https://github.com/vincentmin/privapy}} (\bibinfo{year}{2024}).
\newblock \bibinfo{note}{[Accessed 08-05-2024]}.

\bibitem{master-python-anonymity}
\bibinfo{author}{Madrazo~Quintana, E.}
\newblock \emph{\bibinfo{title}{Building a Python library for anonymizing sensitive data}}.
\newblock \bibinfo{type}{Master's thesis}, \bibinfo{school}{University of Cantabria} (\bibinfo{year}{2023}).
\newblock \bibinfo{note}{\url{https://repositorio.unican.es/xmlui/handle/10902/30791}. Supervisors: López García, Álvaro and Sáinz-Pardo Díaz, Judith}.

\bibitem{prasser2015putting}
\bibinfo{author}{Prasser, F.} \& \bibinfo{author}{Kohlmayer, F.}
\newblock \bibinfo{journal}{\bibinfo{title}{Putting statistical disclosure control into practice: The arx data anonymization tool}}.
\newblock {\emph{\JournalTitle{Medical data privacy handbook}}} \bibinfo{pages}{111--148} (\bibinfo{year}{2015}).

\bibitem{arxaas}
\bibinfo{author}{{NAV IT - The Norwegian Labour and Welfare Directorate}}.
\newblock \bibinfo{title}{Github repository: Arxaas}.
\newblock \bibinfo{howpublished}{\url{https://github.com/navikt/arxaas}} (\bibinfo{year}{2024}).
\newblock \bibinfo{note}{[Accessed 08-05-2024]}.

\bibitem{amnesia}
\bibinfo{author}{{OpenAIRE}}.
\newblock \bibinfo{title}{Amnesia anonymization tool}.
\newblock \bibinfo{howpublished}{\url{https://amnesia.openaire.eu/}} (\bibinfo{year}{2024}).
\newblock \bibinfo{note}{[Accessed 14-05-2024]}.

\bibitem{pycanon_paper}
\bibinfo{author}{S{\'a}inz-Pardo~D{\'\i}az, J.} \& \bibinfo{author}{L{\'o}pez~Garc{\'\i}a, {\'A}.}
\newblock \bibinfo{journal}{\bibinfo{title}{A python library to check the level of anonymity of a dataset}}.
\newblock {\emph{\JournalTitle{Scientific Data}}} \textbf{\bibinfo{volume}{9}}, \bibinfo{pages}{785} (\bibinfo{year}{2022}).

\bibitem{misc_adult_2}
\bibinfo{author}{Becker, B.} \& \bibinfo{author}{Kohavi, R.}
\newblock \bibinfo{title}{{Adult}}.
\newblock \bibinfo{howpublished}{UCI Machine Learning Repository} (\bibinfo{year}{1996}).
\newblock \bibinfo{note}{{DOI}: https://doi.org/10.24432/C5XW20}.

\bibitem{sainzpardo2023comparison}
\bibinfo{author}{S{\'a}inz-Pardo~D{\'\i}az, J.} \& \bibinfo{author}{L{\'o}pez~Garc{\'\i}a, {\'A}.}
\newblock \bibinfo{title}{Comparison of machine learning models applied on anonymized data with different techniques}.
\newblock In \emph{\bibinfo{booktitle}{2023 IEEE International Conference on Cyber Security and Resilience (CSR)}}, \bibinfo{pages}{618--623} (\bibinfo{organization}{IEEE}, \bibinfo{year}{2023}).

\bibitem{poetry}
\bibinfo{author}{{Python-poetry}}.
\newblock \bibinfo{title}{Poetry: Python packaging and dependency management made easy}.
\newblock \bibinfo{howpublished}{\url{https://github.com/python-poetry/poetry}} (\bibinfo{year}{2024}).
\newblock \bibinfo{note}{[Accessed 09-05-2024]}.

\bibitem{pytest}
\bibinfo{author}{{Pytest-dev}}.
\newblock \bibinfo{title}{Github repository: pytest}.
\newblock \bibinfo{howpublished}{\url{https://github.com/pytest-dev/pytest}} (\bibinfo{year}{2024}).
\newblock \bibinfo{note}{[Accessed 14-05-2024]}.

\bibitem{codecov}
\bibinfo{author}{{Sentry}}.
\newblock \bibinfo{title}{Codecov}.
\newblock \bibinfo{howpublished}{\url{https://about.codecov.io/}} (\bibinfo{year}{2024}).
\newblock \bibinfo{note}{[Accessed 15-05-2024]}.

\bibitem{release_please}
\bibinfo{author}{Google}.
\newblock \bibinfo{title}{Release-please project}.
\newblock \bibinfo{howpublished}{\url{https://github.com/googleapis/release-please}} (\bibinfo{year}{2024}).
\newblock \bibinfo{note}{[Accessed 07-05-2024]}.

\bibitem{brandl2021sphinx}
\bibinfo{author}{Brandl, G.}
\newblock \bibinfo{title}{Sphinx documentation}.
\newblock \bibinfo{howpublished}{\url{https://www.sphinx-doc.org/}}.
\newblock \bibinfo{note}{[Accessed 15-05-2024]}.

\bibitem{10.5555/1593511}
\bibinfo{author}{Van~Rossum, G.} \& \bibinfo{author}{Drake, F.~L.}
\newblock \emph{\bibinfo{title}{Python 3 Reference Manual}} (\bibinfo{publisher}{CreateSpace}, \bibinfo{address}{Scotts Valley, CA}, \bibinfo{year}{2009}).

\bibitem{harris2020array}
\bibinfo{author}{Harris, C.~R.} \emph{et~al.}
\newblock \bibinfo{journal}{\bibinfo{title}{Array programming with {NumPy}}}.
\newblock {\emph{\JournalTitle{Nature}}} \textbf{\bibinfo{volume}{585}}, \bibinfo{pages}{357--362}, \url{10.1038/s41586-020-2649-2} (\bibinfo{year}{2020}).

\bibitem{mckinney-proc-scipy-2010}
\bibinfo{author}{{W}es {M}c{K}inney}.
\newblock \bibinfo{title}{{D}ata {S}tructures for {S}tatistical {C}omputing in {P}ython}.
\newblock In \bibinfo{editor}{{S}t\'efan van~der {W}alt} \& \bibinfo{editor}{{J}arrod {M}illman} (eds.) \emph{\bibinfo{booktitle}{{P}roceedings of the 9th {P}ython in {S}cience {C}onference}}, \bibinfo{pages}{56 -- 61}, \url{10.25080/Majora-92bf1922-00a} (\bibinfo{year}{2010}).

\bibitem{typing_extensions}
\bibinfo{author}{{Python}}.
\newblock \bibinfo{title}{Github repository: typing\_extensions}.
\newblock \bibinfo{howpublished}{\url{https://github.com/python/typing_extensions}} (\bibinfo{year}{2024}).
\newblock \bibinfo{note}{[Accessed 15-05-2024]}.

\bibitem{beartype}
\bibinfo{author}{{Beartype developers}}.
\newblock \bibinfo{title}{Github repository: beartype}.
\newblock \bibinfo{howpublished}{\url{https://github.com/beartype/beartype}} (\bibinfo{year}{2024}).
\newblock \bibinfo{note}{[Accessed 15-05-2024]}.

\bibitem{docutils}
\bibinfo{author}{{Docutils developers}}.
\newblock \bibinfo{title}{Docutils: Documentation utilities (webpage)}.
\newblock \bibinfo{howpublished}{\url{https://docutils.sourceforge.io/}} (\bibinfo{year}{2024}).
\newblock \bibinfo{note}{[Accessed 15-05-2024]}.

\end{thebibliography}

\section*{Acknowledgements} 
The authors would like to thank the funding through the project AI4EOSC ``Artificial Intelligence for the European Open Science Cloud'' that has received funding from the European Union's Horizon Europe research and innovation programme under grant agreement number 101058593 and from the SIESTA project ``Secure Interactive Environments for Sensitive daTa Analytics'', funded by the European Union (Horizon Europe) under grant agreement number 101131957.

\section*{Author contributions statement}
Both authors defined and conceived this work. J.S-P.D. performed the formal analysis, methodology, validation and software development. A.L.G. was responsible for supervision, project administration and funding acquisition. Both authors contributed to writing and reviewing the manuscript. 

\section*{Competing interests} 

The authors declare that they have no known competing financial interests or personal relationships that could have appeared to influence the work reported in this paper.

\end{document}